\documentclass[aip,cha]{revtex4-1}

\usepackage{dcolumn}
\usepackage{amsmath}
\usepackage{amssymb}
\usepackage{graphicx}
\usepackage{bm}   
\usepackage{verbatim}
\usepackage{color}
\usepackage{amsthm}

\newcommand{\bra}[1]{\ensuremath{\langle#1|}}
\newcommand{\ket}[1]{\ensuremath{|#1\rangle}}
\newcommand{\bracket}[2]{\ensuremath{\langle#1|#2\rangle}}

\newcommand{\Future}{\stackrel{\rightarrow}{X}}

\newcommand{\Past}{\stackrel{\leftarrow}{X}}

\newcommand{\past}{\stackrel{\leftarrow}{x}}
\newcommand{\State}{\mathcal S}
\newcommand{\Cmu}{C_\mu}

\begin{document}

\title{Increasing complexity with quantum physics}

\author{Janet Anders}
\email{j.anders@ucl.ac.uk}
\affiliation{Department of Physics and Astronomy, University College London, W1E6 BT London, U.K.}
\author{Karoline Wiesner}
\email{k.wiesner@bristol.ac.uk}
\affiliation{School of Mathematics, Centre for Complexity Sciences, University
of Bristol, BS8 1TW Bristol, U.K.}

\bibliographystyle{unsrt}

\begin{abstract}

We argue that complex systems science and the rules of quantum physics are intricately related. We discuss a range of quantum phenomena, such as cryptography, computation and quantum phases, and the rules responsible for their complexity. We identify correlations as a central concept connecting quantum information and complex systems science. We present two examples for the power of correlations: using quantum resources to simulate the correlations of a stochastic process and to implement a classically impossible computational task. 

\end{abstract}

\maketitle

\section*{Lead Paragraph}

A common notion of a complex system is many members acting together to produce an orchestrated and structured ensemble.  Hence, an important feature of a complex system is that it contains a high amount of correlations. Typical examples are neurons in the brain, molecules in a living cell, individual ants in a colony, people in a social network, and the natural forces that together produce climate. Complexity in the regime where quantum mechanics dictates the rules is of a different kind. It arises not because of the sheer number of participants but because the underlying rules of logic change. The Heisenberg principle, for instance, forbids the perfect knowledge of a particle's position and momentum. In this paper we survey how the rules of quantum mechanics lead to a number of unexpected phenomena, such as unbreakably secure communication, quantum parallel computing, and new states of matter, such as a superfluid. Similar to complexity science, in this wealth of quantum phenomena correlations play a central role.
The central role of correlations in both quantum computation and complex systems has practical consequences: To capture all correlations of a complex system in a compact description, it turns out, quantum physics can be very useful.

\section{Introduction}

In complex systems it is often the sheer number of components which is considered to be responsible for the emergence of ``complexity''. However, there is another perspective to how complexity can arise. When the underlying rules of the game change, new structures and phenomena can arise. This is the situation encountered in quantum physics, where quantum objects obey a logic very different from our intuition stemming from the classical world. Quantum laws such as superposition, entanglement and Heisenberg's uncertainty principle imply configurations and relationships that go well beyond what is usually called �complexity�. And they do so already with two constituent parts. 

The aim of this paper is to inform  researchers in complex systems about the intimate relation between quantum information and complexity science. It is not intended as a comprehensive review. Rather its aim is to communicate the aspects of the research on quantum systems produced over the last two decades which is relevant to complex systems science. We give a selective account of a multitude of situations where quantum many-body systems acquire a higher ``complexity'' than classical systems. We then draw the connection between the two fields by focusing on  the importance of correlations for both quantum information and complex systems science. While in complex systems it is intricate (classical) correlations between individual members that are responsible for coherent phenomena such as structure formation and concerted behaviour, quantum complexity relies on quantum correlations that arise as a result of quantum rules. 

The paper is structured as follows. We will first give a brief overview over the relevant quantum phenomena of superposition and entanglement.  We will then present a selection of phenomena showing overwhelming evidence that quantum mechanical correlations are responsible for new, classically unseen phenomena such as unbreakably secure communication, quantum phases, and enhanced computational complexity. We then present two examples that illustrate the relationship between computation, correlation and complexity. The first example shows that structure, i.e. classical correlations, can be generated with fewer computational resources when quantum resources are available. The second example discusses how a limited computer can be enabled to calculate computational task beyond its own capability by accessing quantum correlated resources. Finally we discuss the central role of correlations in both quantum information and complex systems science. 

\section{Laws of quantum physics} \label{sec:laws}

\subsection{Superposition}

The superposition principle lies at the heart of quantum mechanics (see for example \cite{nielsen}). Formally, it is rooted in the linearity of the Hilbert space. Any quantum state can be represented as a vector in Hilbert space. The superposition principle states that any sum of such vectors is also an admissible quantum state. We use the standard \emph{bra, ket} notation, where a bra ($\bra{v}$) is a column vector and a ket ($\ket{v^\prime}$) is a row vector. Hence, $\bracket{v}{v^\prime}$ denotes the inner product between $v$ and $v^\prime$. In other words, if the kets $\ket{\psi_i}$ are quantum states then the superposition
\begin{align}
\ket{\psi} = \sum_i c_i \ket{\psi_i}
\end{align}
is also a quantum state. The $c_i$ here are complex coefficients with the normalisation condition $\sum_i c_i^* c_i^{} = 1$. For a two-state system, any such superposed state is called a \emph{qubit} and its most general form is
\begin{align} \label{eq:super}
\ket{\psi} = \cos(\theta) \ket{0} + e^{i\phi} \sin(\theta) \ket{1}~,
\end{align}
where $\theta$ and $\phi$ are real numbers representing the complex coefficient $c$.
It is important to distinguish between such a \emph{coherent} superposition and a purely classical probability distribution over states of which a given system really only occupies one and we just don't know which. Such a classical probability distribution of quantum states $\ket{\psi_i}$ is formalised by a density operator $\rho$ whose form is independent of the basis of representation
\begin{align}
\rho = \sum_i p_i  \ket{\psi_i}\bra{\psi_i}~,
\end{align}
with $p_i \geq 0$ and $\sum_i p_i = 1$. A so-called \emph{pure} state is a trivial mixture of only one quantum state $\rho = \ket{\psi}\bra{\psi}$. The \emph{von Neumann entropy} of a density operator is defined as
\begin{align}
S(\rho) = - {\rm tr} (\rho \log_2 \rho)~,
\end{align}
where $\rm tr$ is the trace of the matrix. For any pure state $S(\rho) = 0$.

\subsection{Entanglement}

Entanglement is a key feature of quantum mechanics (for a review see \cite{horodecki}). It is the most prominent kind of quantum correlation. However, there are many others such as \emph{discord} \cite{discord}, \emph{quantum conditional entropy} \cite{cond-entr} and minimum entanglement potential \cite{MEP}.
Entanglement is a property assigned to two or more subsystems. Assume that we have two subsystems $S_1$ and $S_2$ and a state vector $\ket{\psi}$ describing the whole system. We say that the subsystems are \emph{entangled} if $\ket{\psi}$ cannot be written as a tensor product of two states $\ket{\psi} = \ket{\phi_1} \otimes \ket{\phi_2}$ where $\ket{\phi_1}$ and $\ket{\phi_2}$ are the states of the  subsystems $S_1$ and $S_2$, respectively. As an example consider what is commonly called a \emph{Bell state}:
\begin{align} \label{eq:Bell}
\ket{\psi_B} = \frac{1}{\sqrt{2}} \left( \ket{0}_1 \otimes \ket{0}_2 + \ket{1}_1\otimes \ket{1}_2  \right)
\end{align}
which is a superposition of two two-dimensional quantum states where the subscript denotes the subsystem. $\ket{\psi_B}$ cannot be written as a tensor product of two separate states for the subsystems $S_1$ and $S_2$. Consequently it is not possible to derive the properties of  $\ket{\psi_B}$ from the sum of the properties of the individual subsystems. 
$\ket{\psi_B}$ contains correlations between $S_1$ and $S_2$ which go beyond what is classically possible. This is expressed in the so-called CHSH inequality \cite{CHSH}. Consider a Bell state which is physically separated and $S_1$ is sent to Alice and subsystem $S_2$ is sent to Bob. Each can then choose one of two bases in which to measure their subsystem in, $A$ and $a$ for Alice, and $B$ and $b$ for Bob. Each local measurement has binary outcomes, either $0$ or $1$. A measure for the correlations of their shared system before the measurement (i.e. of the correlations in $\ket{\psi_B}$) is given by 
\begin{eqnarray} \label{eq:CHSH}
	C 
	&=& 2 (p^s_{AB}+p^s_{Ab}+p^s_{aB}-p^s_{ab}-1)
\end{eqnarray}
where $p^s_{ij}$ is the probability that both, Alice and Bob, will get the same outcome when they measure $i$ and $j$, respectively. The probabilities are normalised such that $p^s_{ij} + p^d_{ij} = 1$, where  $p^d_{ij}$ is the probability that the outcomes of the local measurements are different. If Alice and Bob had classical states, the amount of correlation between the measurement outcomes is upper bounded by $|C_{\rm classical}| \le 2$. However, for quantum states there is a higher bound, $|C_{\rm quantum}| \le 2 \sqrt{2}$. The Bell state reaches the maximum of correlations $|C_{\rm quantum}| = 2 \sqrt{2}$ known as the Tsirelson bound \cite{Tsirelson}.

\subsection{Heisenberg Uncertainty Principle}

Heisenberg stated his uncertainty principle originally as follows: If you make a measurement of an objects's momentum with precision $\Delta p$ you cannot at the same time determine its position with a precision more accurately than $\Delta x = {h \over  4\pi \Delta p}$, where $h$ is Planck's constant. 

Consider the well-known double-slit experiment. An electron is emitted by an electron gun and sent through a wall with two parallel slits in it. Behind the wall there is a fluorescing screen which detects the impact of the electron. After many repetitions of this experiment the screen will show an interference pattern of electron impacts. This is just what is predicted by the superposition principle: The electron's path consists of two superposed paths, one which goes through the left and one which goes through the right slit. Now we move the wall with a well-known speed perpendicular to the two slits. The impact of the electron onto either the left slit or the right slit will change the momentum of the wall and hence, without measuring the electron, just by measuring the change in momentum of the wall, we can determine which of the two paths the electron has actually taken. This contradicts the superposition principle of two perfectly simultaneous realisations in one system. However, this contradiction is solved when Heisenberg's uncertainty principle is taken into account. Since we cannot measure the momentum of the wall with perfect accuracy while also determining its position with the same accuracy we cannot know exactly where the two slits are. This is enough to smear out the interference pattern of the electrons. Hence, it is a general rule of quantum mechanics that one cannot design equipment that  determines which of two alternatives in a superposition is realised without, at the same time, destroying the superposition \cite{feynman}.

\section{Quantum complexity}

Given the rules of quantum physics we may expect new kinds of complexity at the atomic level. Here we will give an overview of a range of phenomena where quantum resources lead to the emergence of a higher complexity. 

\subsection{Quantum cryptography}

Cryptography is the science of encoding and transmitting a secret message in such a way that it is very hard for an eavesdropper to break the code and learn the message. This problem can be reduced to transmitting a secret key, such as a sequence of \emph{random} binary numbers with which a binary message is encrypted by adding the two, e.g. random key $ = 0011001$, message to be encoded $= 1110000$ and cipher $= 1101001$. The randomness of the key then guarantees the security of the encoded message. However, this still means that a \emph{secret key} has to be shared in the first place between two communicating parties, traditionally called Alice and Bob. Today many cryptographic techniques that achieve this, so called key distribution schemes, rely on mathematical functions that behave like a trap-door: it is easy to get in, but hard to get out. An example is the public key RSA scheme that relies on the difficulty of factoring. Here the easy direction of the trap door is the simplicity of multiplying two (rather large) prime numbers. The difficulty lies in the reverse direction - finding the two unknown prime factors of a large number. In fact there is no known classical algorithm that can factor in any reasonable time  (polynomial time), and the factoring problem lies in a computational complexity class  (NP) that is conjectured to contain problems that require exponential time to solve \cite{factoring, NP}. However, the bottom line is that the security of RSA, and all classical algorithms that use mathematical trap door functions, relies solely on the \emph{assumption} that reversing the trap door is computationally hard, i.e. it takes exponentially long to solve. 

This is why the excitement was big when in 1984 Brassard, a computer scientist, and Bennett, a physicist, realised that quantum physics allows the sharing of a provably secure key \cite{BB84}. They developed the first quantum key distribution scheme, the BB84 protocol, in which the sender, Alice, prepares her key in a qubit. For this she chooses one of two bases at random, the computational basis  $\{|0 \rangle, |1 \rangle\}$ or the superposed $x$-basis $\{|+\rangle = {|0 \rangle + |1 \rangle \over \sqrt{2}}, |-\rangle = {|0 \rangle - |1 \rangle \over \sqrt{2}}\}$. She then sends the qubit to her communication partner, Bob, who will measure in the same two bases, also at random. In the classical world an intercepting eavesdropper wanting to learn the key would take a copy of the transmitted information and then attempt to decrypt the message. Copying an unknown qubit, however, is not permitted. This is known as the \emph{no-cloning principle}, a theorem that results from Heisenberg's uncertainty principle. So an eavesdropper is faced with the task of intercepting the actual message carrier, attempt to decrypt it, and pass the resulting qubit - or a totally new qubit - on to Bob. To successfully decrypt the eavesdropper has to guess in which basis, the computational or the superposition basis, Alice used to encode the qubit in. This has two effects. Firstly a measurement in the ``correct'' basis will reveal perfectly which state was encoded, while measuring in the ``wrong basis'' results in no information for the eavesdropper. This is because the $x$-basis corresponds to equal probabilites to find $|0 \rangle$ or $|1 \rangle$, and vice versa. The second effect is that when the eavesdropper measures in the ``wrong'' basis the state changes. This is another consequence of the Heisenberg uncertainty principle and it is this second feature which makes BB84 unbreakably secure. More precisely,  the communicating parties are able to monitor how much an eavesdropper could have learnt by sifting for only those bits where they happened to use the same basis. They can then identify if and how much the  information sent was tinkered with. This knowledge allows Alice and Bob to ensure that the amount of leaked information to the eavesdropper about their key is arbitrarily small (through a classical scheme, called privacy amplification). To summarise, BB84 achieves a feat no classical cryptographic scheme can dream of - a provably secure key distribution protocol. This security relies solely on the laws of quantum physics, in particular, the superposition principle and Heisenberg's uncertainty principle  \footnote{It has been shown that the security can be formulated in a device-independent way, just relying on the quantum rules outlined in section \ref{sec:laws} \cite{Acin}. However, it should also be noted that the cryptographic security of a scheme is as weak as its weakest link and attacks on the authenticity of the signal have been reported \cite{Norway}.}.


\subsection{Quantum computing} \label{sec:computing}

It was a quirk of (quantum) history, that in 1994 Shor, a mathematician, showed that factoring, the mathematical function safeguarding the widely used RSA protocol, can be solved in polynomial time on a quantum computer \cite{shor}. Shor's algorithm not only made it clear that the security of governments, banks and private persons was at  a very real risk if a quantum computer was built, it also implied that the complexity of computational problems could change significantly when quantum rules come into play. 
To clarify what constitutes a quantum computer, the straightforward way is to think of a classical computer with some input bits and a series of gates, such as XOR, NOT and AND, that are applied to produce the output of the computation. This setup can be ``quantised'', i.e. instead of the classical states, $0$ or $1$, the input for a quantum computer is the superposition state of Eq.~(\ref{eq:super}). A reversible gate set, consisting, for instance, of Pauli gates, Hadamard, and  CNOT gates \cite{reversible-gates},  can then coherently manipulate the superposition states. This intrinsic parallel capability can be identified as the fundamental reason for the quantum computer to lead to the immense speed up over (known) classical algorithms. Simply speaking, instead of one computation, starting with say a single input $0$, the quantum computer runs two computations, with both inputs $0$ and $1$, in parallel.

However, there are other, equivalent, models of quantum computation where other quantum effects lie behind the change of computational complexity. Known models include adiabatic quantum computation and topological quantum computation \cite{AQC, TQC}. One radically different way of performing a quantum computation, and one that lacks a classical analogy, is measurement-based quantum computation (MBQC) \cite{MBQC}. Here computation relies on the use of a computational resource state, a highly \emph{entangled} multi-qubit state. Many resource states can be represented by graphs. The graph vertices symbolise qubits and the graph edges indicate that entangling CNOT operations were applied to the connected vertices. A computation in MBQC is achieved by applying a sequence of adaptive measurements on individual qubits. Due to the underlying correlations between the qubits this allows information to be processed. The adaptiveness of measurements further guarantees that even though measurements are inherently random the overall output of the computation is deterministic. The ``resource'' for the exponential speed up of the MBQC model of quantum computation lies in the presence of entanglement in the resource state. 

It is clear that (still theoretical) quantum computers allow an exponential speed up over (known) classical algorithms of solving certain computational problems. The reason for this increase in computational complexity can be traced to quantum rules, such as superposition and especially entanglement, in different computational models. However, it remains unresolved whether the exponential speed up is a truly quantum feature in the sense that there exists no classical algorithm that could achieve a solution exponentially quickly  \cite{NP}. In contrast, we will present a computational situation in subsection \ref{subsec:computabilityincrease} where quantum devices lead to not just a speed up but a provable increase of the actual ability to compute a certain task.

\subsection{Quantum phases}

Standard phase transitions appear when the temperature of a material, such as water, is varied across a critical value. Thermal fluctuations in the material then cause the formation of a new phase, such as ice. However, in classical systems, when absolute zero temperature is reached all thermal fluctuations are frozen out and the material is fixed in its state even when other parameters are changed, such as the pressure. In contrast, a quantum system shows fluctuations even at zero temperature as a consequence of Heisenberg's uncertainty principle. Now these quantum fluctuations can drive phase transitions, between quantum phases, even at zero temperature.  One example is the \emph{quantum phase transition} from a superfluid to a Mott insulator, first observed in the lab in 2002 \cite{Greiner02}. A three-dimensional optical lattice is created  with local trapping potentials of depth $U$ at each of the $M$ lattice points. $N$ quantum particles, i.e. Rubidium atoms, are then loaded into the lattice and allowed to hop between neighbouring sites with a probability proportional to $J$. An important quantum characteristic is here that the atoms are Bosons, that is, they follow Bose-Einstein statistics instead of classical Boltzmannian statistics. Again this is a consequence of the non-commutativity of quantum mechanics and thus related to Heisenberg's uncertainty principle. In the limit of weak interaction $U >> J$, the trapped atomic gas is in the Mott insulator phase, with its quantum state being given by 
\begin{equation}
	| \psi_{MI} \rangle \propto \bigotimes_{j=1}^M \left(a_j^{\dag}\right)^n|0\rangle. 
\end{equation}
Here $| 0 \rangle$ is the vacuum state of the lattice on which $n = N/M$ creation operators, $(a_j^{\dag})^n$ are applied, each filling a Boson in the individual lattice site, $j$. Since the atoms are pin-pointed to their lattice sites and do not move,  they can not help a current to flow, and the material is an insulator. However, when the trapping potential is reduced and crosses a \emph{critical value}, a new phase is formed. For $J >> U$ the gas of atoms behaves as a superfluid. This is because the atoms are shared coherently, in superposition, across the whole lattice, as expressed in the superfluid state 
\begin{equation}
	| \psi_{SF} \rangle \propto \left( \sum_{j=1}^M {a_j^{\dag} \over \sqrt{M}} \right)^N |0\rangle,
\end{equation}
where each single Boson is coherently spread, in superposition, over the whole lattice,  ${a_1^{\dag} + a_2^{\dag} + ... + a_M^{\dag} \over \sqrt{M}}$. The superfluid phase is a genuinely quantum phase with exceptional characteristics, in particular, it flows without any friction. Quantum rules thus open up a plethora of qualitatively new arrangements of materials that would be impossible in a classical world. For a wide range of important physical phases it is now clear that quantum entanglement plays a central role. One example is the formation of a truly quantum phase, the Bose-Einstein condensate, another is the BCS theory of superconductivity \cite{Anders-BEC, Anders-phases}.

\subsection{Quantum effects in biology and thermodynamics} 

Quantum mechanics is crucial for our understanding of the stability of atoms, the periodic table of chemical elements, and determining how molecules are formed. However, when it comes to macroscopic organisms, such as cats and humans, our reasonable assumption is that all quantum effects vanish at this scale, due to decoherence \cite{Zurek}. Yet exactly where  the crossover is where quantum mechanics ceases to be relevant is a topic of intense debate. Excitingly, biological structures, such as the helical structure of DNA, and biological functions such as photosynthesis and the navigation mechanism that allows birds to find their way using the earth magnetic field have all been linked to quantum effects \cite{Rieper-DNA, photosynthesis, bird-nav}. In photosynthesis, for example,  energy in form of photons from the sun is captured by chromophores in bacteria or the leaves of plants. These photons are converted into electronic excitations on a molecular level and are transported to a so-called reaction centre. 
Curiously, this transport happens with a much higher efficiency than what classical models predict. Now quantum transport models are being applied to this process and show good agreement \cite{photosynthesis}. The ability of organisms to make use of absorbed sun light would then rely on the ability of a quantum excitation to be shared between different parts of the cell, in superposition, just like in the quantum superfluid phase. But it is not just biology at the microscopic scale where quantum effects bring a new twist. Even on macroscopic scales, where statistical mechanics and thermodynamics describe the behaviour, quantum mechanics could lead to qualitatively different phenomena. Even the laws of thermodynamics have been questioned and modified \cite{Hilt-Landauer} when applied to quantum systems. Revising thermodynamics in the light of the development of quantum mechanics over the last century is an emerging field where some questions on the foundations of statistical mechanics have already been resolved \cite{Winter-Short}.

\section{Complexity and computation} \label{sec:examples}

Choosing correlations as the starting point the following two examples illustrate the intricate relationship between complexity and computation. 
The first example is concerned with the number of bits needed to fully characterise  the structure of a given system, i.e. all its correlations. It turns out that this requires fewer qubits than classical bits.

A key question in the field of computational complexity is what tasks a given system can perform. The second example is concerned with a particular computational task and illustrates that quantum correlations are necessary to solve it under certain constraints.

\subsection{Generating correlations more efficiently with quantum resources}\label{subsec:stochasticprocess}

A stochastic process is a toy model of a complex system. Being discrete and one-dimensional, it is well suited to analyse the formation of structure, a key feature of a complex system. Stochastic processes have been used to study structure formation in for example self-organisation, protein dynamics, neural dynamics \cite{shalizi_quantifying_2004, li_multiscale_2008, haslinger_computational_2010}). 

For our purposes, a stochastic process is a probability distribution $\Pr(\Past;\Future)$ over a bi-infinite sequence of random variables $X$. A random variable $X$ is a probability distribution $\Pr(X=x)$ over an alphabet $x \in {\mathcal X}$, the letters in the alphabet are the possible observations. We assume stationarity, i.e. $\Pr(X_{t_0}^{t_{n-1}}=x^n) = \Pr(X_{t_{0+k}}^{t_{n-1+k}}=x^n)$. $\Future$ denotes the observations of the process after some reference time $t_0$ and $\Past$ are the observations up to $t_0$. \emph{Structure} is simply the sum of all correlations in such a sequence of observations. 

The resources required to generate a stochastic process -- that is to realise the probability distribution $\Pr(\Past;\Future)$ -- are measured by the minimal size of a representation storing all the information about the correlations \cite{crutchfield_inferring_1989,shalizi_computational_2001}. Such a representation needs to distinguish between differing sequences of observations only if their conditional future observations statistically differ. This yields an equivalence relation $\sim$ between all past observations, such that $\past\sim\past^\prime$ iff $\Pr(\Future | \Past = \past) = \Pr(\Future | \Past = \past^\prime)$. Grouping together equivalent observations into \emph{equivalence classes} $s_i \equiv \epsilon(\past) = \{\past^\prime : \past\sim\past^\prime\}$ we obtain a provably  minimal representation of the correlations as a list of the equivalence classes ${\State}$, and a transition function between them $T_{ij}^x = \Pr( X=x, \State=s_j | {\State} = s_i)$ \cite{shalizi_computational_2001}\footnote{In statistics, this is known as a \emph{sufficient statistic}.}. The entropy of the probability distribution over these states is a measure of the resources required to generate the stochastic process:
\begin{align}
\Cmu \equiv H({\State}) = -\sum_{i} \Pr(s_i) \log \Pr(s_i)~.
\end{align}
$\Cmu$ is called the \emph{statistical complexity} and is measured in bits \cite{crutchfield_inferring_1989}.

To break the classical limit of $\Cmu$, we construct the following quantum finite-state representation. The basis set of the quantum states are constructed from the classical states $\State_i$ and the letters in the alphabet $x$. Hence, a measurement of such a state will yield a new state and an output symbol. This way, the original stochastic process is generated through successive measurements \cite{gu_sharpening_2011}:
\begin{align}
\label{eq.qStates}
\ket{\psi_j} = \sum_{\State_i \in {\bf S}} \sum_{x\in {\mathcal X}} \sqrt{T_{ij}^x}\ket{\State_i}\ket{x}~.
\end{align}
 
The resulting mixed state $\rho = \sum_{j } \Pr(\State_j) \ket{\psi_j}\bra{\psi_j}$ has an entropy of 
\begin{align}
\label{eq.Cq}
C_q \equiv S(\rho) = - {\rm tr} \rho \log \rho~,
\end{align}

$C_q$ is measured in qubits. It was shown that $C_q \leq \Cmu$, and in most cases $C_q$ is strictly less than $\Cmu$ \cite{gu_sharpening_2011}. In other words, to generate the same amount of correlations requires fewer quantum resources than classical resources. The relevant resources are the classical and quantum states, respectively, which are the equivalent to the number of bits and quantum bits, respectively, which need to be stored by some computational device. If, for example, $\Cmu$ is just above $2$ bits and $C_q$ just under $2$ qubits it means that the classical implementation requires $3$ bits where the quantum implementation only requires $2$ qubits, as we will see in the example in Section~\ref{subsec:stochasticprocess}.

\begin{figure}[htbp]
\begin{center}
	{\includegraphics[width=0.4\textwidth]{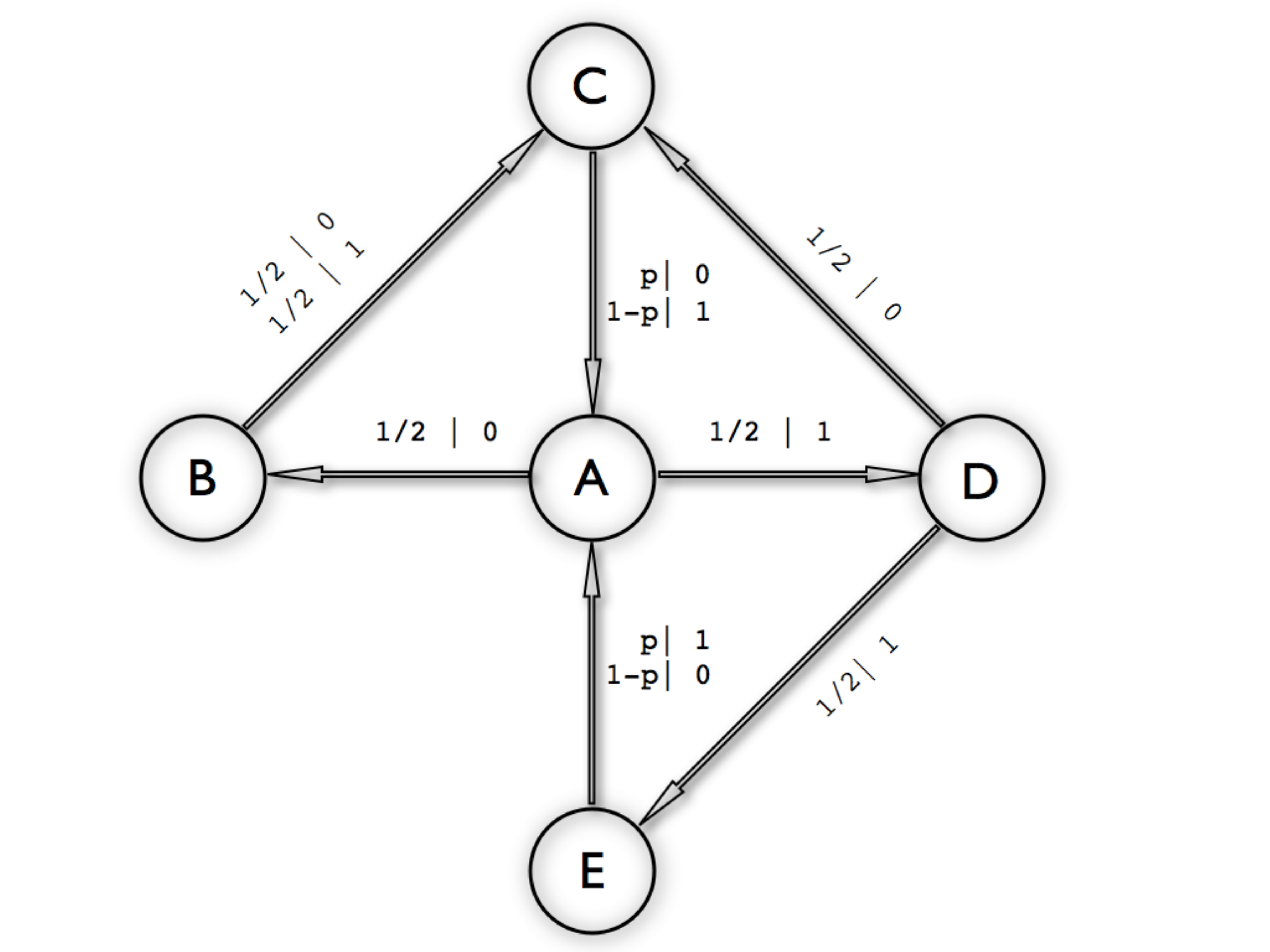}}
\caption{Equivalence classes for the AND process. States represent equivalence classes, edges are labeled with transition probabilities and output symbols ($T_{ij}^x$). Going from state to state according to the probabilities on the edges will generate the AND process. $p$ ($1-p$) is the probability of outputting the result of a logical AND (NAND) on the last two output bits.}
\label{fig.machine}
\end{center}
\end{figure}

We illustrate this with the following example of a stochastic process. Consider an infinite sequence of concatenated blocks of three symbols where the first two are random and the third symbol is the logical AND of the first two (see Table~\ref{tab:multiply}). Later on we will add a stochastic element. The alphabet is ${\cal X} = \{0,1\}$. There are five equivalence classes for this process, labeled $A-E$ and represented as circles or \emph{states} in Fig.~\ref{fig.machine}. The $T_{ij}^x$ label the \emph{edges} between the states (setting $p=1$ for now). We can compute the state probabilities from the left eigenvector of the transition matrix $\sum_{x\in\cal X} [T_{ij}^x]$ and obtain for the statistical complexity $\Cmu = 2.19$ bits. Hence, the generation of these correlations requires classical resources of at least $2.19$ bits. 

Constructing the quantum states using Eq.~\ref{eq.qStates} we obtain the following set of states
\begin{table}[h]
\begin{tabular}{l  l}
\ket{\psi_1} = & $\frac{1}{\sqrt{2}} (\ket{B}\otimes\ket{0} + \ket{D}\otimes\ket{1})$ \\
\ket{\psi_2} = & $\frac{1}{\sqrt{2}} (\ket{C}\otimes\ket{0} + \ket{C}\otimes\ket{1}) $\\
\ket{\psi_3} = & $\ket{A}\otimes\ket{0}$\\
\ket{\psi_4} = & $\frac{1}{\sqrt{2}} (\ket{C}\otimes\ket{0} + \ket{E}\otimes\ket{1})$ \\
\ket{\psi_5} = & $\ket{A}\otimes\ket{1}$\\
\end{tabular}
\end{table}

Computing the quantum and classical complexity for the process we obtain $\Cmu = 2.19$ bits and $C_q = 2.13$ qubits. So, the amount of quantum resources required to simulate this stochastic process is slightly lower than classically. Now, we introduce an element of stochasticity into the logical operation. The probability of computing AND will now be $p$ and the probability of computing NAND will be $1-p$. This is equivalent to a {\it noisy} AND gate. The $T_{ij}^x$ change accordingly, see Fig.~\ref{fig.machine}. Two of the quantum states change to the following states
\begin{table}[h]
\begin{tabular}{l  l}
\ket{\psi_3} = & $\sqrt{p}\ket{A}\otimes\ket{0} + \sqrt{1-p}\ket{A}\otimes\ket{1}$ \\
\ket{\psi_5} = & $\sqrt{p}\ket{A}\otimes\ket{1} + \sqrt{1-p}\ket{A}\otimes\ket{0}$ \\
\end{tabular}
\end{table}

\begin{figure}[htbp]
\begin{center}
	{\includegraphics[width=0.5\textwidth]{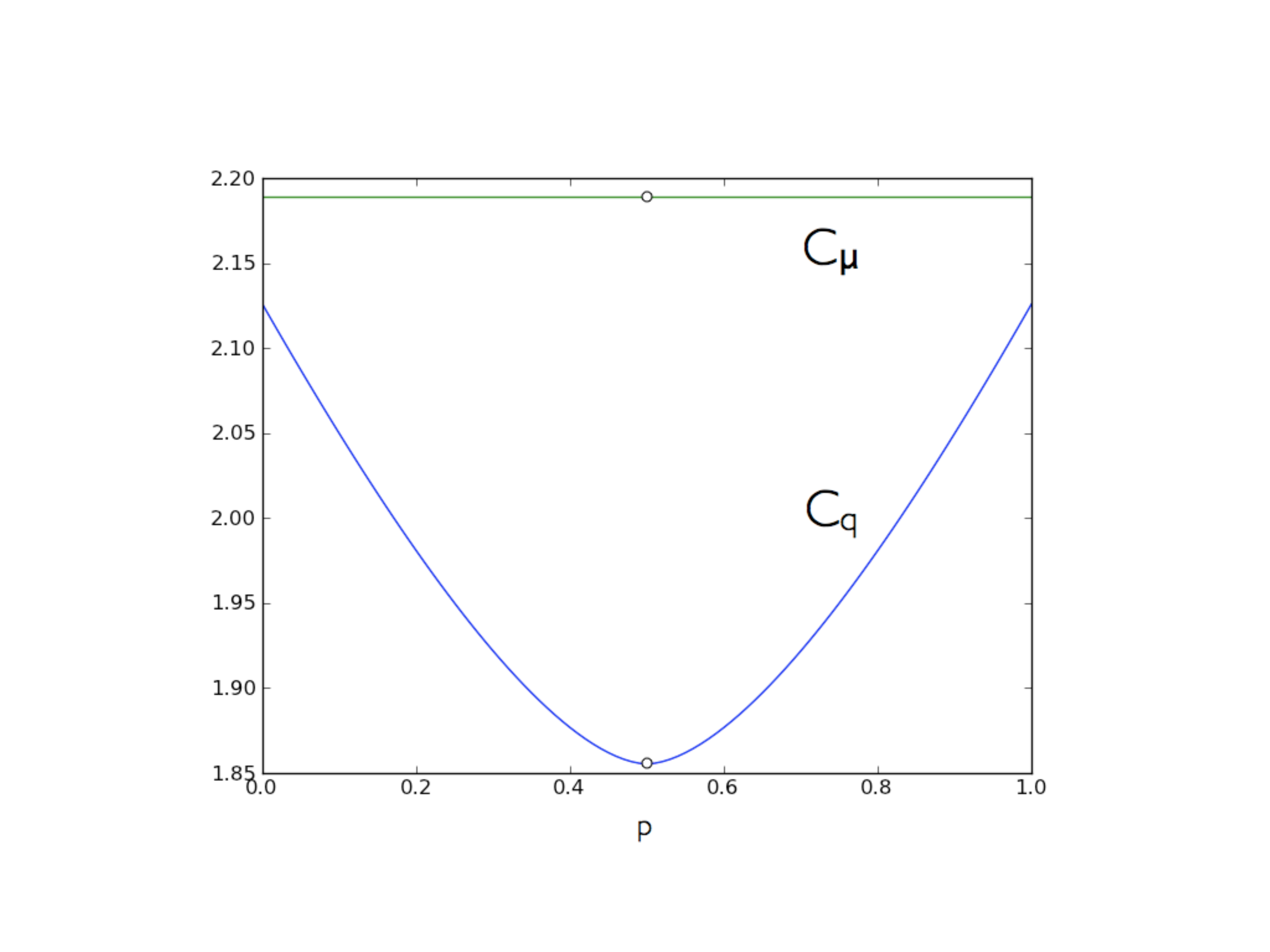}}
\caption{$\Cmu$ and $C_q$ for the AND process. (For $p=1/2$ both are zero.)}
\label{fig.CmuCq}
\end{center}
\end{figure}

Computing $\Cmu$ and $C_q$ anew we notice that the $\Cmu$ is independent of $p$. In other words, although we are generating a process with fewer correlations we still need the same amount of classical resources (only for $p=1/2$ $C_q = \Cmu = 0$). $C_q$, however, drops steadily as $p$ increases from $0$ to $1/2$. So, indeed, the fewer correlations we want to generate the fewer resources we require. $\Cmu$ and $C_q$ are plotted in Fig.~\ref{fig.CmuCq}.

The reason that $C_q$ changes as a function of $p$ is the ability of quantum states to be in a superposition of varying degree. For increasing $p$ the distinction between states $\ket{\psi_3}$ and $\ket{\psi_5}$ becomes less and less relevant and we can put them closer and closer to each other in Hilbert space. Classically we can't do this. There is no analogue to a quantum superposition in classical physics.

There are processes for which there is no difference in the number of required bits and qubits, $\Cmu = C_q$. In \cite{gu_sharpening_2011} it was shown that this is the case if and only if the generation using the equivalence classes is reversible in the following sense: Given the current state and observation the previous state is determined \cite{wiesner_information_2009}. Indeed, if we choose the XOR logical operation (see Table~\ref{tab:add}) instead of the AND operation we find that $\Cmu = C_q$ for any $p$ and no gain with quantum resources is possible. 
We will see a similar (ir)reversibility in the next example.

Recently, evidence started emerging that some biological processes contain a quantum component. Most prominent are suggestions that photosynthesis is such an example \cite{engel_evidence_2007, lee_coherence_2007}. There are still many open questions. The above framework provides tools for analysing classical and a quantum aspects of information processing in a stochastic process from a rather general starting point -- a data sequence only. Although the example is constructed from logic gates it illustrates the advantage of quantum over classical resources for a stochastic process. The following will illustrate the other side of the same coin: What role do quantum correlations play for the power of simple classical and quantum logic operations.

\subsection{Raising the computational complexity using quantum correlations} \label{subsec:computabilityincrease}

In Section \ref{sec:computing} we saw that quantum computation promises algorithms that solve a range of problems exponentially faster than known classical algorithms. However, since no proof excludes the possibility of an equally fast classical algorithm the quantum speed up remains a (likely) conjecture \footnote{A provably faster than classical quantum algorithm is the Grover algorithm. The speed up is quadratic, not exponential.}.  In contrast we will now discuss an example where quantum correlations provide truly new computational power, not just a speed up, to solve an otherwise impossible task. 

Suppose we are given two classical bits of information, the inputs $a, b \in \{0, 1\}$ and a very limited (and therefore totally old fashioned) pocket calculator. This pocket calculator can only do the following operations: output a constant, output the input, and output the binary sum, the XOR, of two inputs, see Tab.~\ref{tab:add}. 
\begin{table}[h]
\begin{tabular}{|c|c|c|} 
\hline
$a$ & $b$ & $a\oplus b$ \\
\hline
0 & 0 & 0 \\
0 & 1 & 1 \\
1 & 0 & 1 \\
1 & 1 & 0 \\
\hline
\end{tabular}
\caption{ \label{tab:add} Truth table for the binary sum $(\oplus)$, also called  XOR gate. (Its reversible version with outputs $a$ and $a \oplus b$ is known as the CNOT gate.) }
\end{table}
Now our task is to calculate the product of the two inputs, $a \otimes b$, also known as the AND gate, see Tab.~\ref{tab:multiply}, with the help of our pocket computer.
\begin{table}[h]
\begin{tabular}{|c|c|c|}
\hline
$a$ & $b$ & $a\otimes b$ \\
\hline
0 & 0 & 0 \\
0 & 1 & 0 \\
1 & 0 & 0 \\
1 & 1 & 1 \\
\hline
\end{tabular}
\caption{\label{tab:multiply} Truth table for product $(\otimes)$, also called AND gate. }
\end{table}
We notice that binary addition is a \emph{balanced} or \emph{linear} Boolean function, which is invertible when one of the inputs is kept as an output. In contrast, the product is an \emph{unbalanced}, or \emph{non-linear}, Boolean function and hence not invertible. It is easy to prove that any combination of balanced functions can never result in an unbalanced function. We thus conclude it is impossible for our pocket computer to perform the required multiplication. 

\begin{figure}[b]
\medskip
   \begin{center}
	\includegraphics[width=0.055\textwidth]{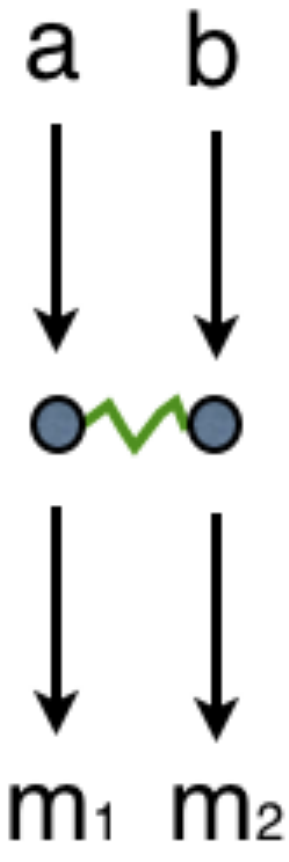} \quad \quad \quad
	\includegraphics[width=0.094\textwidth]{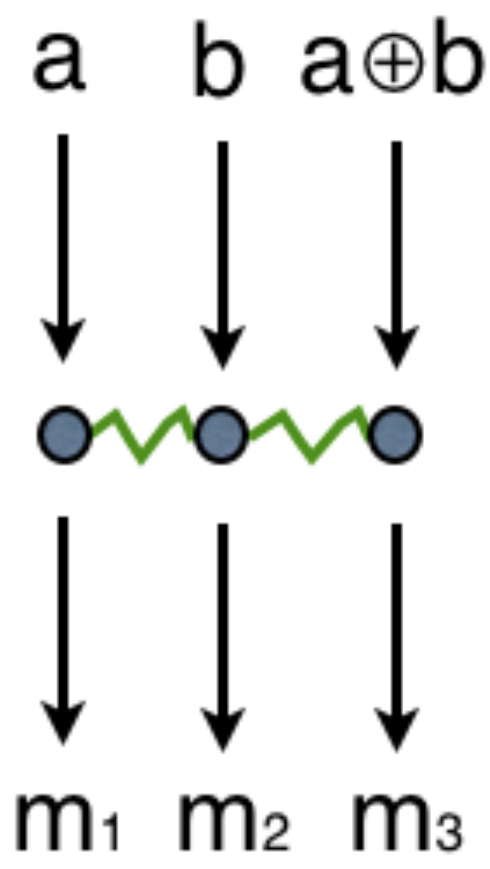}
        \caption{\label{fig:sites} Left panel: Two sites (circles) with shared correlations (zig-zag line) receive the inputs $a$ and $b$ and output $m_1$ and $m_2$, respectively. Right panel: Three correlated sites, receiving inputs $a, b$ and $a \oplus b$, respectively.} 
   \end{center}
\end{figure}

From the discussion of quantum cryptography, computation and phases we took the insight that quantum correlations can lead to a substantial qualitative change. Inspired by these examples one may wonder if the impossible task would become feasible by allowing our pocket computer to access \emph{correlated sites}. To answer this let us consider two sites, as depicted in the left panel in Fig. \ref{fig:sites}, that may share correlations between them, either classical or quantum. 
The type of shared correlations is not specified here as we want to determine what sort of correlations we \emph{require} to solve the computational task. We use the pocket computer to send inputs to the correlated sites and receive outputs, as depicted in Fig. \ref{fig:sites}. For two sites, the best option is to input the two bits we want to multiply, $a$ and $b$, each to one of the two sites. As a consequence of the sites' previously shared correlations the returned outputs, $m_1$ and $m_2$, may now be correlated too. The possibility that emerges here is that the correlations of the outputs could be such that their binary sum is just the required product of the inputs, $m_1 \oplus m_2 = a \otimes b$. Importantly, this sum is computable with the pocket computer. So all we need is to find the correct correlations between the sites that produce the required correlated outputs. This can be analysed in terms of the average probability of success of this strategy, 
\begin{eqnarray}
P_{\rm av. succ.} &=& {1 \over 4} \, \sum_{a, b \in \{0, 1 \}} p_{\rm succ}(m_1 \oplus m_2 = a \otimes b),
\end{eqnarray}
where $p_{\rm succ}(m_1 \oplus m_2 = a \otimes b)$ is the probability that the outputs $m_1$ and $m_2$ add to give the product for a specific pair of inputs $a$ and $b$. The average success probability can be rearranged \cite{VanDam05} to bring it into the form of the CHSH correlation measure introduced in Eq.~(\ref{eq:CHSH}), 
\begin{eqnarray}
P_{\rm av. succ.} &=& {p^s_{00}+p^s_{01}+p^s_{10}-p^s_{11}+1  \over 4} \\
		&=& { C  \over 8 } + {1 \over 2 } 
\end{eqnarray}
where $p^s_{ab}$ are the probabilities that given their two respective inputs, $a$ and $b$, the two sites output the same bits, i.e. $m_1 = m_2$, and $C$ quantifies the CHSH correlations between the two sites. The maximum success of calculating $a \otimes b$ thus directly depends on the upper bound of correlations between the two sites. As discussed in section \ref{sec:laws}, classical correlations are bounded by $|C_{\rm classical}| \le 2$ and therefore $P^{\rm classical}_{\rm av. succ.} \le {1 + 2 \over 4 } = 75\% $. This is a trivial result since 75\% is just the success rate of adopting a constant 0 output to predict the multiplication of two arbitrary inputs $a, b$, see Tab. \ref{tab:multiply}. However, when the two sites are quantum correlated the maximum probability grows beyond the trivial benchmark of 75\%, up to $P^{\rm quantum}_{\rm av. succ.} \approx 85\%$ for the maximally entangled Bell state, Eq.~(\ref{eq:Bell}), with $|C_{\rm quantum}| = \sqrt{2} \, 2$.  Moreover, it can be shown that with three sites the success probability for $a \otimes b$ can be brought to 100\% while classical states can never achieve 100\% \cite{Anders-PRL}. The quantum state with the required correlations is the three-party Greenberger-Horne-Zeilinger (GHZ) state \cite{GHZ}, $|\psi_{GHZ} \rangle = { |0_1 0_2 1_3\rangle + |1_1 1_2 0_3\rangle \over \sqrt{2}} $, a famous entangled state \cite{Mermin}. To make the result explicit, the inputs $a$, $b$ and $a \oplus b$ will determine measurements on the three sites of the state, see right panel in Fig. \ref{fig:sites}. Two different measurement bases are used depending on the input bit. For input  0 the measurement basis is the  $x$-basis $\{|+ \rangle, |-\rangle \}$ while for input 1 the measurement basis is the $y$-basis $\{|+i\rangle = {|0 \rangle + i | 1 \rangle \over \sqrt{2}} , |-i \rangle = {|0 \rangle - i | 1 \rangle \over \sqrt{2}}\}$. 
The output bits are $0$ for a ``$+$'' outcome and $1$ for a ``$-$'' outcome. In conclusion, we have found that a limited XOR computer can be boosted to compute incomputable functions, such as AND, when given access to quantum correlations. While the exponential speed up of quantum computers over classical computers remains a conjecture, the increase in computability due to quantum correlations is a provable fact clearly exposing the computational advantage of quantum correlations over classical ones. 

\subsection{Discussion of complexity and correlations}
\begin{figure}
\medskip
   \begin{center}
	\includegraphics[width=0.55\textwidth]{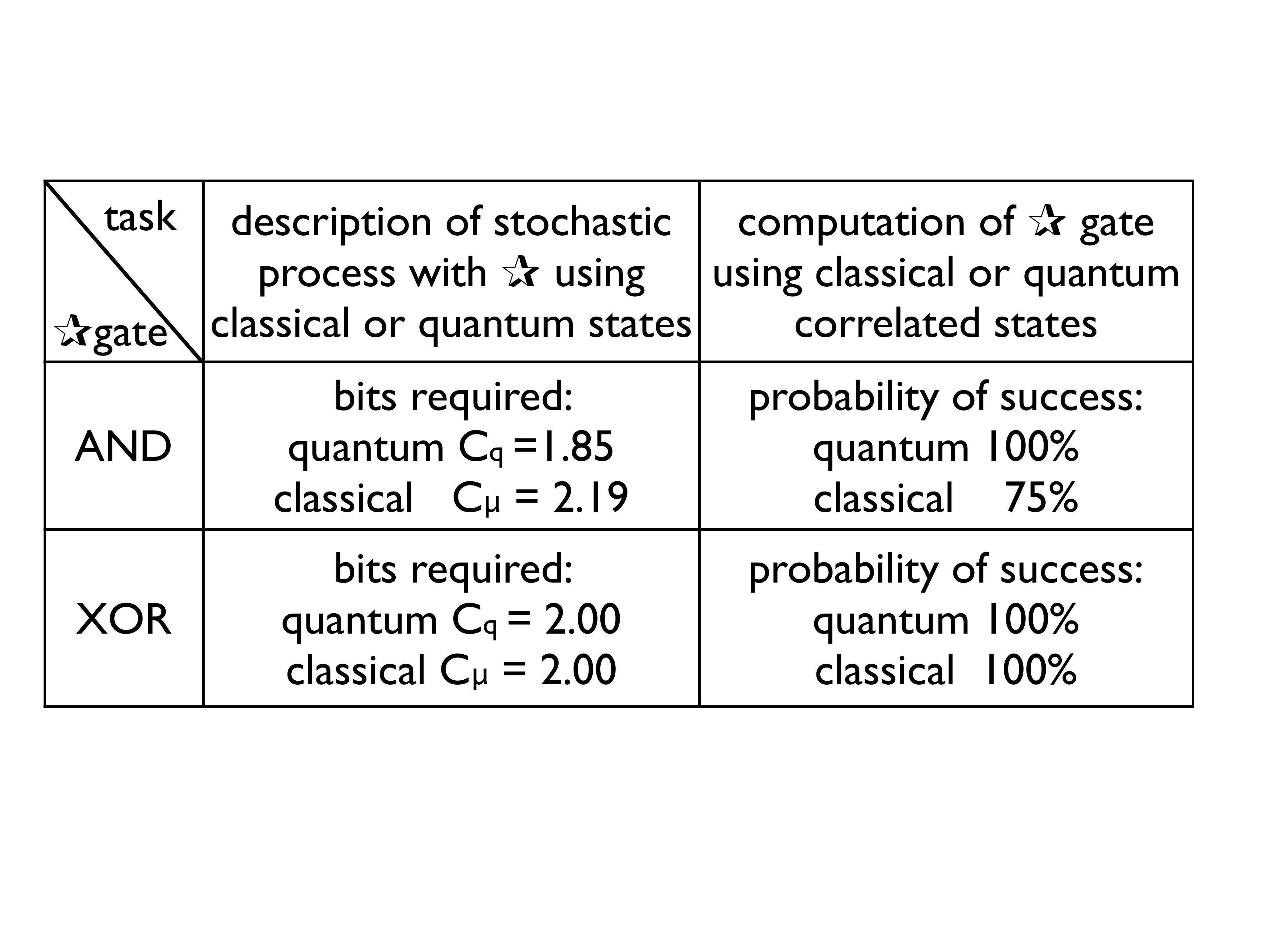}
        \caption{\label{fig:table} A summary of the two examples. Left column: The stochastic process involving the AND logical operation is more efficiently simulated using quantum resources. Right column: The computational task of an AND logical gate can only be performed with 100\% accuracy using quantum correlations (under certain constraints). For details see text.} 
   \end{center}
\end{figure}

In both examples we consider the logical operations AND and XOR. Figure~\ref{fig:table} summarises the commonalities. In Section~\ref{subsec:computabilityincrease} we saw that under certain constraints the implementation of the AND gate can only be done error free if one has access to quantum correlations. The implementation of the XOR gate, on the other hand, is unaffected by access to quantum resources. The reason for this lies in the linear vs non-linear character of the XOR and AND operation, respectively, an intriguing subject unfortunately outside of the scope of this article. The example in Section~\ref{subsec:stochasticprocess} gave a complimentary perspective on the power of correlations. Here, it was the simulation of correlations which could be done with fewer computational resources when quantum physics was used. Here, too, this advantage was only present for a process involving the AND logical operation and not for the XOR operation. This is for the same reasons as above, the non-linearity of the AND operation leads to this gain in efficiency.

\section{Discussion}

We have seen that quantum physics offers a wide range of new phenomena, including unbreakably secure cryptography, exponentially fast computations, the superfluid phase, and even fast exciton transport in photosynthesis.  In the field of complex systems  such qualitatively new properties that {\it emerge} out of the constituent parts and their interactions are considered a feature of  \emph{complexity}.  The striking effect of quantum resources in all of the above examples is that they assist in increasing the complexity. A high amount of correlations between constituent parts is also a feature of a complex system, and often used to quantify its complexity. In the field of quantum information entanglement, a well-defined and particularly striking type of (quantum) correlations, has turned out to be of central importance for exactly that reason. It effectively measures the complexity of quantum systems and processes.  However, entanglement is not the only indicator of complexity, weaker quantum correlations exist, such as discord, quantum relative entropy and minimum entanglement potential \cite{discord, cond-entr,MEP}. They have in common that they are all purely quantum features and their presence can enhance quantum computations \cite{Nature-discord}.

In the light of the above we believe that complex systems science, despite the different physics, can learn from the central role quantum correlations play for such `emergent' phenomena. 
Taking correlations as our starting point we have exemplified that 
 two originally separate meanings of the term complexity can be united. On the one hand one speaks of complexity in a complex system which represents the amount of structure, among other things. On the other hand we speak of computational complexity as a measure of difficulty of solving a computational task. The examples illustrated that the use of the word {\it complexity} in both cases is not a coincidence but rather a signature of the commonalities between the two research areas. Both are concerned with the power of correlations.

{\bf Acknowledgements} 
JA acknowledges funding from the Royal Society. KW acknowledges funding through EPSRC grant EP/E501214/1 and the Santa Fe Institute for their hospitality during the workshop.


\begin{thebibliography}{99}


\bibitem{nielsen} M. A. Nielsen and I. L. Chuang, Quantum Computation and Quantum Information. Cambridge University Press, 2000.

\bibitem{feynman} R. P. Feynman, R. B. Leighton, and M. Sands, Lectures on Physics: Quantum Mechanics v. 3, 1st ed. Addison Wesley, 1971.


\bibitem{Anders-PRL}
	J. Anders, D. E. Browne,
	Phys. Rev. Lett. 102, 050502 (2009). 

\bibitem{Nature-discord}
	Z. Merali,
	Nature 474, 24 (2011).

\bibitem{discord}
	H. Ollivier and W.H. Zurek,
	Phys. Rev. Lett. 88, 017901 (2001).
	
\bibitem{cond-entr}
	M. Horodecki, J. Oppenheim and A. Winter,
	Nature 436, 673 (2005)
	
\bibitem{MEP}
	M. Piani, S. Gharibian, G. Adesso, J. Calsamiglia, P. Horodecki, A. Winter, 
	Phys. Rev. Lett. 106, 220403 (2011). 


\bibitem{Tsirelson} 
	B. S. Tsirelson, 
	Lett. Math. Phys. 4, 93 (1980).

\bibitem{factoring}
Factoring is a function problem and actually in the complexity class FNP, { \verb| http://en.wikipedia.org/wiki/Integer_factorization |}, a sister of the more famous NP \cite{NP}. 

\bibitem{NP}
The computational complexity class NP has appeared in the popular media recently, {\verb| http://www.newscientist.com/article/dn19287-p |} \\ {\verb| --np-its-bad-news-for-the-power-of-computing.html|}, both, because of its central importance for computer science and quantum computing, and because of its prize money of 1 Million USD, {\verb| http://www.claymath.org/millennium/P_vs_NP/|}. 

\bibitem{BB84} 
	C. H. Bennett and G. Brassard, 
	in Proceedings of IEEE Intern. Conf. on Computer Systems and Signal Processing (1984), Volume: 11, 
	IEEE Press , 175 (1984).

\bibitem{reversible-gates} 
	C. H. Bennett,  
	IBM Journal of Research and Development 17, 525 (1973).

\bibitem{AQC}
	E. Farhi, J. Goldstone, S. Gutmann, M. Sipser,
	arXiv:quant-ph/0001106v1 (2000).
	
\bibitem{TQC}
	First proposed by A. Kitaev;
	S. Das Sarma, M. Freedman and C. Nayak, 
	Phys. Rev. Lett. 94, 166802 (2005).
	
\bibitem{shor} P. W. Shor, 
	SIAM J.SCI.STATIST.COMPUT. 26, 1484 (1997).
	arXiv:quant-ph/9508027v2 (1995).


\bibitem{MBQC}
	R. Raussendorf and H. J. Briegel, 
	Phys. Rev. Lett. 86, 5188 (2001); 
	R. Raussendorf, D. E. Browne, and H. J. Briegel, 
	Phys. Rev. A 68, 022312 (2003);
	H. J. Briegel, D. E. Browne, W. D\"ur, R. Raussendorf and M. Van den Nest, 
	Nature Physics 5, 19 (2009).

\bibitem{MBQC=PT} 
	J. Anders, M. Hajdusek, D. Markham, V. Vedral, 
	Foundations of Physics 38, 506 (2008). 
	
\bibitem{Greiner02} 
	M. Greiner, O. Mandel, T. Esslinger, T. H\"ansch, I. Bloch, 
	Nature 415, 39 (2002).

\bibitem{CHSH}
J. F. Clauser, M. A. Horne, A. Shimony, and R. A. Holt, 
	Phys. Rev. Lett. 23, 880 (1969). 

\bibitem{GHZ}
	D. M. Greenberger, M. A. Horne, and A. Zeilinger, in Bell�s Theorem, Quantum Theory, and Conceptions, edited by M. Kafatos (Kluwer Academic, Dordrecht, 1989), pp. 69.
	
\bibitem{Mermin} 
	N. D. Mermin, Am. J. Phys. 58, 731 (1990). 
	
\bibitem{Acin}
	A. Ac\'in, N. Brunner, N. Gisin, S. Massar, S. Pironio, V. Scarani,
	Phys. Rev. Lett. 98, 230501 (2007).

\bibitem{Norway}
	L. Lydersen, et al., 
	Nature Photonics {\bf 4} 686 (2010).
	
\bibitem{Anders-phases}
	J. Anders, V. Vedral, 
	Open Systems \& Information Dynamics 14, 1 (2007);
	E. Rieper, J. Anders, V. Vedral, 
	New Journal Physics 12, 025017 (2010).

\bibitem{Anders-BEC}
	J. Anders, D. Kaszlikowski, Ch. Lunkes, T. Ohshima, V. Vedral, 
	New Journal of Physics 8, 140 (2006).


\bibitem{Zurek}
	W. H. Zurek, 
	Reviews of Modern Physics 75, 715 (2003).
	
\bibitem{Rieper-DNA}
	E. Rieper, J. Anders, V. Vedral,
	arXiv:1006.4053v2 (2010).
	
\bibitem{photosynthesis}
	M. Mohseni, P. Rebentrost, S. Lloyd, and A. Aspuru-Guzik,
	The Journal of Chemical Physics, 129(17):174106, (2008); 
	F. Caruso, A. W. Chin, A. Datta, S. F. Huelga, and M. B. Plenio,
	The Journal of Chemical Physics, 131(10):105106, (2009); 
	F. Fassioli, A. Olaya-Castro, S. Scheuring, J. N. Sturgis, and N.F. Johnson,
	Biophysical journal, 97(9):2464�2473, 11 (2009). 

\bibitem{bird-nav} 
	E. M. Gauger, E. Rieper, J. J. L. Morton, S.C. Benjamin, and V. Vedral,
	Phys. Rev. Lett. 106, 040503 (2011);
	J. Cai, G.G. Guerreschi, and H.J. Briegel,
	Phys. Rev. Lett. 104, 220502 (2010).

\bibitem{Hilt-Landauer}
	S. Hilt, S. Shabbir, J. Anders, E. Lutz, 
	Physical Review E (Rapid), 83, 030102 (2011).

\bibitem{Winter-Short}
	S. Popescu, A. J. Short, A. Winter, 
	Nature Physics 2, 754 (2006).

\bibitem{VanDam05}
W. van Dam, arXiv: quant-ph/0501159 (2005).
M. Hoban, E. Campbell, K. Loukopoulos, D.E. Browne,
New Journal of Physics 13, 023014 (2011).

\bibitem{crutchfield_inferring_1989} J. P. Crutchfield and K. Young, 
	Phys. Rev. Lett. 63, 105 (1989).

\bibitem{shalizi_computational_2001} C. R. Shalizi and J. P. Crutchfield, 
	J. Stat. Phys. 104, 817 (2001).

\bibitem{gu_sharpening_2011} M. Gu, K. Wiesner, E. Rieper, and V. Vedral, 
	arXiv:1102.1994 (2011).

\bibitem{wiesner_information_2009} K. Wiesner, M. Gu, E. Rieper, and V. Vedral, 
	 arXiv:0905.2918 (2009).

\bibitem{li_multiscale_2008} C.-B. Li, H. Yang, and T. Komatsuzaki, 
	Proc. Nat. Acad. Sci. 105, 536 (2008).

\bibitem{shalizi_quantifying_2004} C. R. Shalizi, K. L. Shalizi, and R. Haslinger, 
	Phys. Rev. Lett. 93, 118701 (2004).

\bibitem{haslinger_computational_2010} R. Haslinger, K. L. Klinkner, and C. R. Shalizi, 
	Neural Computation 22, 121 (2010)

\bibitem{engel_evidence_2007} G. S. Engel et al., 
	 Nature 446, 782 (2007).

\bibitem{lee_coherence_2007} H. Lee, Y.-C. Cheng, and G. R. Fleming, 
	Science 316, 1462 (2007).
 

\bibitem{horodecki}
	R. Horodecki, P. Horodecki, M. Horodecki, K. Horodecki,
	 Rev. Mod. Phys. 81, 865 (2009).


\end{thebibliography}
\end{document}